# Core-Polarization and Relativistic Effects in Electron Affinity Calculations for Atoms: A Complex Angular Momentum Investigation

Z. Felfli and A.Z. Msezane
Center for Theoretical Studies of Physical Systems,
Clark Atlanta University, Atlanta, GA 30314 USA

Core-polarization interactions are investigated in low-energy electron elastic scattering from the atoms In, Sn, Eu, Au and At through the calculation of their electron affinities (EAs). The complex angular momentum (CAM) method wherein is embedded the vital electron-electron correlations is used. The core-polarization effects are studied through the well investigated rational function approximation of the Thomas-Fermi potential, which can be analytically continued into the complex plane. The EAs are extracted from the large resonance peaks in the CAM calculated low-energy electron – atom scattering total cross sections and compared with those from measurements and sophisticated theoretical methods. It is concluded that when the electron-electron correlation effects and core polarization interactions are accounted for adequately the importance of relativity on the calculation of the EAs of atoms can be assessed. For the At atom relativistic effects are estimated to contribute a maximum of about 3.6% to its EA calculation.

PACS number(s): 34.80.Bm, 32.10.Hq

## I. Introduction

Essential to understanding the physics of cooling and trapping of gaseous atomic ensembles and of cold plasmas, including new molecules creation from atoms, is the knowledge of low-energy collisions of atoms and ions [1]. Negative ion formation through low-energy collisions as intermediate resonances provides deep insight into quantum dynamics [2] and defines the mechanism through which electrons deposit energy and induce chemical transitions [3]. The polarization potential of the neutral state causes attachment of a weakly bound electron, producing inter-shell-type resonances [4]. The vital importance of the core-polarization interaction in low-energy electron scattering from atoms and molecules has been demonstrated [4-9]. Ground state formation of a negative ion is identifiable through the large resonance peak in the low-energy electron – atom scattering total cross section (TCS) [10]. This facilitates the extraction of the EAs from the electron elastic scattering TCSs.

Electron–electron correlations and core-polarization interactions are crucial for the existence and stability of most negative ions. The extent of importance of relativistic effects in the calculation of the electron affinities (EAs) of atoms has, to our knowledge, not been carefully assessed. Here we use the complex angular momentum (CAM), also known as the Regge pole methodology [11] wherein is embedded the electron–electron correlations, to investigate core-polarization interactions. The well investigated rational function approximation of the Thomas-Fermi (T-F) potential [12-14] is adopted; it was obtained from the Thomas-Fermi theory of the atom [15-17]. It is noted that the important Ramsauer–Townsend (R-T) minima that characterize low-energy electron elastic scattering total cross sections manifest the polarization of the atomic core by the scattered electrons [18]. Relativistic effects are assessed in the determination of the EAs of the representative atoms: In, Sn, Eu, Au and At with Z values of 49, 50, 63, 79 and 85, respectively. This is accomplished by comparing the CAM calculated EAs with those from various measurements and sophisticated relativistic calculations. The electron affinity calculations of atoms probe the many-body effects in many electron systems; therefore, their calculation provides a stringent test of theoretical methods when the results are compared with those from reliable measurements.

The main motivation for the present investigation is the recent photodetachment measurements of the EA of the open d- and f-subshell Ce atom (Z=58) [19-21]. These measurements were aided by the multi configurational Dirac-Fock-relativistic configuration interaction (MCDF-RCI) calculations [22, 23]. As pointed out in [19] the determination of the adiabatic EA of Ce via the photodetachment transition is a challenge. The recently measured EAs [19-21] of Ce are very close to those calculated using the MCDF-RCI [22, 23], Relativistic energy-consistent small-core lanthanide pseudopotential multireference configuration-interaction (MRCI) [24], and Regge pole [25] methods. An extensive comparison of the EAs of Ce obtained by various experiments and calculations is found in [25]. This is indicative of the extent of the significance of relativistic effects in the calculation of the EA of the Ce atom. Consequently, by controlling the electron-electron correlation effects and core polarization interactions, the importance of the relativistic effects in the calculations of the EAs of atoms can be assessed.

**Atomic In**

The measured EA of In [26] has been used in [27] to benchmark the CAM methodology and validate the effectiveness of the T-F type potential for use in intermediate Z atoms. The power and beauty of the CAM method is that it requires no external input data, either experimental or theoretical; it extracts the EA from the resonances in the electron elastic scattering TCS. The agreement between the CAM calculated and measured [26] EA of In was found to be excellent. The CAM calculated EA has also been compared with other values [28-35].

**Atomic Sn**

The selection of the Sn atom is dictated by the availability of extensive accurate EAs, both experimental, measured mainly by laser photodetachment electron spectrometry (LPES) and laser-photodetachment threshold (LPT) spectroscopy, and theoretical [36-42]. The most recently measured value is by laser photodetachment microscopy (LPM) [43]. These EAs for Sn and the measured EA for In using infrared photodetachment threshold spectroscopy [26] agree excellently with our CAM calculated values. It is noted that in our calculated elastic TCSs for the ground state In and Sn atoms the bound states of the In and Sn negative ions appear as isolated resonances. This facilitated the unambiguous identification and extraction of the EAs of In and Sn in their elastic TCSs.

**Atomic Eu**

With a relatively high Z of 63, but small EA the Eu atom provides a stringent test of the CAM method when its prediction (EA=0.116eV) [25] is compared with that calculated using the MCDF-RCI (EA=0.117 eV). [44]. The measured EA for Eu is 1.053 eV [45], which is about a factor of 10 larger than the calculated values. Suffice to state that the main objective of this paper is to investigate core-polarization interaction on the calculation of the EA and assess the importance/unimportance of relativistic effects. So, the experimental value is only informative.

**Atomic Au**

We have selected the Au atom because: 1) Several measurements and calculations agree on the large value of its EA [46-51]; 2) Nanogold is a good catalyst, but bulk Au is not [52]; 3) The $Au^-$ anion is important in catalyzing the oxidation of $H_2O$ to $H_2O_2$ through the anionic molecular complex $Au^-(H_2O)_{1,2}$ formation in the transition state [53] and the oxidation of $CH_4$ to methanol without $CO_2$ emission [54]. We stress that the CAM method extracts the EA of Au and other atoms from the dramatically sharp peak in the electron elastic scattering TCS, consistent with the experimental prescription in [10]. The success of the nonrelativistic CAM theory which embeds the vital electron-electron correlation effects has been attributed mainly to the accurate description of the crucial core-polarization interaction. This does not contradict that relativistic effects are important in gold chemistry in general, Gorin and Toste [55] and Hakkinen et al. [56] and references therein.

Accounting for relativistic effects does not necessarily ensure reliable calculated EAs if the crucial electron-electron correlations and core-polarization interactions have not been adequately accounted for. Many existing binding energies of atomic negative ions are calculated using structure-based theoretical methods. Therefore the EAs calculated by these methods are often riddled with uncertainty and lack definitiveness for complex systems. For instance, Relativistic effects in gold chemistry were investigated by Wesendrup et al [50] who performed large scale fully relativistic Dirac–Hartree–Fock and MP2 benchmark calculations as well as Nonrelativistic pseudopotential calculations and obtained the EAs of 2.19 eV and 1.17 eV, respectively. These values should be contrasted with the CAM calculated value of 2.263 eV [49] and compared with the measured EAs [46-48]

**Atomic At**

For the At atom there are no experimental EAs available to our knowledge. However, long ago, Zollweg [57], using a semiempirical extrapolation method, calculated the EA of At to be 2.80 eV. More recently, the CAM method [27] and the multiconfiguration Dirac –Hartree-Fock (MCDHF) method [36] calculated the EA to be 2.51 eV and 2.421 eV, respectively. These calculated values are sufficient for our objective.

The structure of the paper is as follows: Section II presents the method of calculation. In Section III is given the results, while in Section IV the summary and conclusions are presented.

## II. Method of Calculation

Regge poles are singularities of the S matrix. Therefore, they rigorously define resonances [58, 59], thereby making the Regge-pole methodology appropriate for our investigation. Regge trajectories probe electron attachment at the fundamental level near threshold; they penetrate the atomic core, thereby allow the determination of reliable electron affinities of complex atomic and molecular systems. The importance of Regge trajectories in low energy electron scattering has been investigated recently by Thylwe and McCabe [60] using the T-F type potential [12]. The results for the Xe atom are presented in their Fig. 2, where the Dirac Relativistic and non-Relativistic Regge trajectories are contrasted. The investigation found the interesting result, namely when Im L is very small, both calculations yield essentially the same Re L at resonance (L is the complex angular momentum). This is indicative of the insignificance of the difference between the Relativistic and non-Relativistic calculations at low scattering energies, corresponding to electron attachment, leading to negative ion formation as resonances.

Here we calculate the electron elastic TCSs and the Mulholland partial cross sections for only the ground state atoms and extract their EAs. The Im L is used to distinguish between the shape resonances (short-lived resonances) and the stable bound states of the negative ions (long-lived resonances) formed as Regge resonances during the electron-atom collision. For the latter the Im L is several orders-of-magnitude smaller than that for the former (see comparisons in for example Ref. [25]). In the CAM description of scattering we use the Mulholland formula [61] in the form [11] (atomic units are used throughout):

$$\sigma_{tot}(E) = 4\pi k^{-2} \int_0^\infty \mathrm{Re}[1 - S(\lambda)]\lambda d\lambda$$
$$- 8\pi^2 k^{-2} \sum_n \mathrm{Im} \frac{\lambda_n \rho_n}{1 + \exp(-2\pi i \lambda_n)} + I(E) \quad (1)$$

where S is the S-matrix, $k = \sqrt{(2mE)}$, with m being the mass, $\rho_n$ the residue of the S-matrix at the nth pole, $\lambda_n$ and I(E) contains the contributions from the integrals along the imaginary λ-axis; its contribution has been demonstrated to be negligible [25]. We will consider the case for which Im $\lambda_n \ll 1$ so that for constructive addition, $\mathrm{Re}\lambda_n \approx 1/2, 3/2, 5/2...$, yielding $\ell = \mathrm{Re} L \cong 0, 1, 2...$ The significance of Eq. (1) is that a resonance is likely to influence the elastic TCS when its Regge pole position is close to a real integer [11].

The calculation of the elastic TCSs and the Mulholland partial cross sections uses the well investigated rational function approximation of the Thomas-Fermi (T-F) potential [12-14]

$$U(r) = \frac{-Z}{r(1 + aZ^{1/3}r)(1 + bZ^{2/3}r^2)}, \quad (5)$$

where $Z$ is the nuclear charge and *a* and *b* are adjustable parameters. For small r, the potential describes the Coulomb attraction between an electron and a nucleus, $U(r) \sim -Z/(r)$, while at large distances it mimics the polarization potential, $U(r) \sim -1/(abr^4)$ and accounts properly for the vital core-polarization interaction at low energies. The effective potential

$$V(r) = U(r) + L(L+1)/(2r^2), \quad (6)$$

is considered here as a continuous function of the variables $r$ and L. The potential, Eq. (5) has been used successfully with the appropriate values of *a* and *b*. When the TCS as a function of "*b*" has a resonance [11] corresponding to the formation of a stable bound negative ion, this resonance is longest lived for a given value of the energy which corresponds to the EA of the system (for ground state collisions). This was found to be the case for all the systems we have investigated thus far and fixes the optimal value of "*b*" for Eq. (5).

For the numerical evaluation of the TCSs and the Mulholland partial cross sections, we solve the Schrödinger equation for complex values of L and real, positive values of E

$$\psi'' + 2\left(E - \frac{L(L+1)}{2r^2} - U(r)\right)\psi = 0, \tag{7}$$

with the boundary conditions:

$$\psi(0) = 0,$$
$$\psi(r) \sim e^{+i\sqrt{2E}r}, \; r \to \infty. \tag{8}$$

We note that Eq. (8) defines a bound state when $k \equiv \sqrt{(2E)}$ is purely imaginary positive. We calculate the S-matrix, $S(L, k)$ poles positions and residues of Eq. (7) following a method similar to that of Burke and Tate [62]. In the method the two linearly independent solutions, $f_L$ and $g_L$, of the Schrödinger equation are evaluated as Bessel functions of complex order and the S-matrix, which is defined by the asymptotic boundary condition of the solution of the Schrödinger equation, is thus evaluated. Further details of the calculation may be found in [62].

In Connor [63] and Ref. [11] the physical interpretation of Im L is given. It corresponds inversely to the angular life of the complex formed during the collision. A small Im L implies that the system orbits many times before decaying, while a large Im L value denotes a short-lived state. For a true bound state, namely $E < 0$, Im L $\equiv$ 0 and therefore the angular life, 1/[Im L] $\to \infty$, implying that the system can never decay. Im L is also used to differentiate subtleties between the ground and the excited states of the negative ions formed as resonances during the collisions.

### III. Results

In Figs 1 through 5 are presented the variation of the TCSs (a.u) as a function of E (eV) at fixed a=0.2 for various values of the "b" parameter near resonances, corresponding to electron attachment. Figure 1 presents the result for the In atom, showing clearly the characteristic near-threshold behavior of the TCS, namely the appearance of the R-T minimum followed by the shape resonance. For the "b" parameters displayed, the positions of both the R-T minimum and the shape resonance are not sensitive. However, the extreme sensitivity of the calculated EA at 0.380 eV of In on "b" is evident. The range of "b" values 0.0330-0.0328 result in the EA of between 0.560 eV and 0.380 eV, indicative of the strong sensitivity of the calculated EA to the polarization interaction. These EAs can be contrasted with the latest measured value of 0.385 eV [26]. When b changes from 0.0328 to 0.0327, a very small change in b of 0.0001, the resonance disappears, demonstrating that the conditions for In¯ formation no longer hold.

Just like for In there are many measurements of the EA of Sn as well as theoretical calculations. The agreement among the measurements themselves is very good and with the theoretical calculations as well. Here we simply adopt the latest measured value of 1.112070 eV [43] to compare with our CAM calculated value. Figure 2 displays the data for the Sn atom in the energy region of its EA. As in Fig. 1, the characteristic R-T minimum and the shape resonance are clearly captured. Here also both their positions are not sensitive when b changes by 0.0001, namely from 0.0350 to 0.0351. However, the EA changes from about 1.10 eV to 1.15 eV. When b changes from 0.035 to 0.034 the resonance at 1.10 eV disappears, implying that the polarization potential can no longer support a Sn¯ negative ion.

As pointed out in the Introduction the EA of Eu has been determined theoretically to be about 0.117 eV, while the experimental value is 1.053 eV. Since the purpose of the paper is to investigate the effects of core polarization and assess the importance of relativity on the calculation of the EA, it is adequate to compare the CAM calculated value with that obtained by the MCDF-RCI method. Figure 3 displays the results for the Eu atom. In this case the characteristic R-T minimum is at a lower energy than shown in the figure, but the shape resonance is clearly visible and like in the previous cases, it is not much sensitive to the polarization interaction. Also displayed are the curves corresponding to the various values of b in the neighborhood of the EA. A change of 0.0001 in the b parameter, namely from 0.0375 to 0.0376 changes the EA from 0.117 eV to 0.150 eV , demonstrating the sensitivity of the EA to the polarization interaction. Also the excellent agreement between the CAM calculated EA and that from the MCDF-RCI method leads us to conclude that the nonrelativistic CAM method is adequate in calculating the EA of Eu.

The EA of Au has been investigated extensively, both theoretically and experimentally. Experiments agree on the EA of Au; its EA thus provides a stringent test of theoretical methods. Incidentally, the CAM calculated EA agrees excellently with measurements. Relativistic effects are known to be important in gold chemistry, Gorin and Toste [55] and Hakkinen et al [56]. How important are they in the calculation of the EA of Au? Figure 4 displays the Au TCS (a.u) versus E (eV) for various values of the parameter b. The characteristic R-T minimum and the shape resonance appear near threshold as expected in the TCS. Importantly, here a second R-T minimum appears together with the EA of Au at 2.26 eV. The position of the R-T minimum is less sensitive to a small variation in the b parameter compared to the position of the shape resonance. When the b parameter changes from 0.0360 to 0.0370, the EA changes from 2.26 eV to 3.80 eV, indicative of the importance of the polarization interaction on the calculation of the EA of Au. A change of the b parameter from 0.0360 to 0.0358 prevents the attachment process. The appearance of the EA at the second R-T minimum has already been discussed in the context of the Au⁻ anion catalysis [53, 54].

The TCS for At is shown in Fig. 5 for different values of b. Like those of Au, these TCSs are characterized by R-T minimum and shape resonance near threshold. Around the second R-T minimum appears the sharp resonance, identified through Im L as corresponding to the EA of At, determined to be 2.51 eV. Save for its height the position of the first R-T minimum is not sensitive to small variations in b, from 0.043 to 0.042. By comparison the position and height of the shape resonance are moderately sensitive to the above small changes in b. But the value of the EA is very sensitive to small variations in b, changing from 2.51eV to 4.00 eV, when b varies from 0.043 to 0.042. However, the depths of both minima are also sensitive to these small variations in b. When the value of b is 0.041 the polarization potential can no longer support electron attachment.

The At results allow us to assess the extent of relativistic effects on the calculation of the electron affinity for the At atom through comparison with other theoretical calculations such as the relativistic calculated EA by Li *et al.* [36]. Table 1 presents the R-T minima, shape resonances and EAs for the atoms In, Sn, Eu, Au, and At extracted from the TCSs of Figures 1 through 5, respectively. For comparison, the latest experimental and theoretical EAs for the In, Sn and Au atoms are also included. For the Eu and At atoms the EAs are those from the MCDF-RCI [44] and MCDHF [36] calculations, respectively. To our knowledge, there are no experimental or other theoretical R-T minima and shape resonances available for these atoms to compare with. These R-T minima and shape resonances call for both experimental and other theoretical investigations for confirmation.

### IV.    Summary and Conclusions

CAM investigations have been carried out of low-energy electron elastic scattering from the representative atoms In, Sn, Eu, Au and At through the calculation of their TCSs. The objective has been to study the effects of the core-polarization interactions on the calculation of their EAs as well as assess the significance of relativistic effects. The CAM method embodies the vital electron-electron correlation effects while the core-polarization interactions are included in the well investigated rational function approximation of the Thomas-Fermi potential, which can be analytically continued into the complex plane. The sensitivity of the EAs to small variations in the "b" parameter of the core-polarization potential has been demonstrated. The extracted EAs from the TCSs are found to be in excellent agreement with the measured ones for In, Sn and Au and with the MCDF-RCI calculated EA for Eu [44]. For the At atom, there are no measured values available, but the CAM calculated EA agrees very well with the recent value from the multiconfiguration Dirac –Hartree-Fock method [36] and the EA of Zollweg [57].

Also determined are the characteristic Ramsauer-Townsend minima, important in sympathetic cooling and the production of cold molecules from natural fermions, and the shape resonances, useful for interpreting chemical reactions yielding negative ion formation. Most important here is that the EAs for the Au and At atoms appear almost at the second R-T minima of their TCSs, demonstrating the necessary catalytic properties of both the negative $Au^-$ and $At^-$ ions for direct use in the oxidation of $H_2O$ to $H_2O_2$ and $CH_4$ to methanol without $CO_2$ emission. As pointed out in the Introduction, the R-T minima are the essential features manifesting the polarization of the atomic core by the scattered electron [18]. Also, ground state atoms with low EA values are needed for quenching Rydberg atoms through collisions.

From comparison of the calculated EA of the At atom using the MCDHF method [36] and our nonrelativistic CAM method, it is concluded that the contribution from relativity is less than about 3.6%. This is consistent with the conclusion from nanoscale catalysis using negative ions [53], where the calculation of the transition state barriers for $H_2O$, HDO and $D_2O$, when catalyzed by the $Au^-$ ion to the corresponding peroxides, found relativistic effects to contribute only less than 2.3%. Both the all-electron relativistic potential and the non-relativistic potential were used in that assessment. The great advantage of the CAM method over other theoretical methods is that it does not require any experimental or theoretical input data; it simply extracts the EAs from the resonances in the electron elastic TCSs.

### Acknowledgements

This research is dedicated to Professor Alex Dalgarno, FRS for interest in and unwavering and strong support of our research. Research was supported by the US DOE, Division of Chemical Sciences, Office of Basic Energy Sciences, Office of Energy Research and CAU CFNM, NSF-CREST PROGRAM. The research used resources of the National Energy Research Scientific Computing Center, which is supported by the Office of Science of the US DOE under Contract No. DE-AC02-05CH11231

| Atom | Z | b | 1st R-T minimum | SR | 2nd R-T minimum | EA Present | EA Expt. | EA Theory |
|---|---|---|---|---|---|---|---|---|
| In | 49 | 0.0328 | 0.068 | 0.238 | N/A | 0.380 | 0.385[26] 0.409[35] | 0.380[29] 0.393[30] 0.403[33] |
| Sn | 50 | 0.0350 | 0.058 | 0.309 | N/A | 1.100 | 1.112[43] 1.112[40] | 1.128[36] 1.093[42] |
| Eu | 63 | 0.0375 | N/A | 0.027 | N/A | 0.116 | 1.053[45] | 0.117[44] |
| Au | 79 | 0.0360 | 0.69 | 1.07 | 2.06 | 2.26 | 2.309[34] 2.31[47] | 2.263[49] 2.19[50] |
| At | 85 | 0.0420 | 0.82 | 1.26 | 2.40 | 2.51 | N/A | 2.80[57] 2.421[36] |

**Table 1:** Ramsauer–Townsend (R-T) minima, Shape Resonances (SRs), Electron Affinities (EAs), all in eV, and optimized b parameters of the polarization potential for the atoms In, Sn, Eu, Au, and At. N/A represents not available.

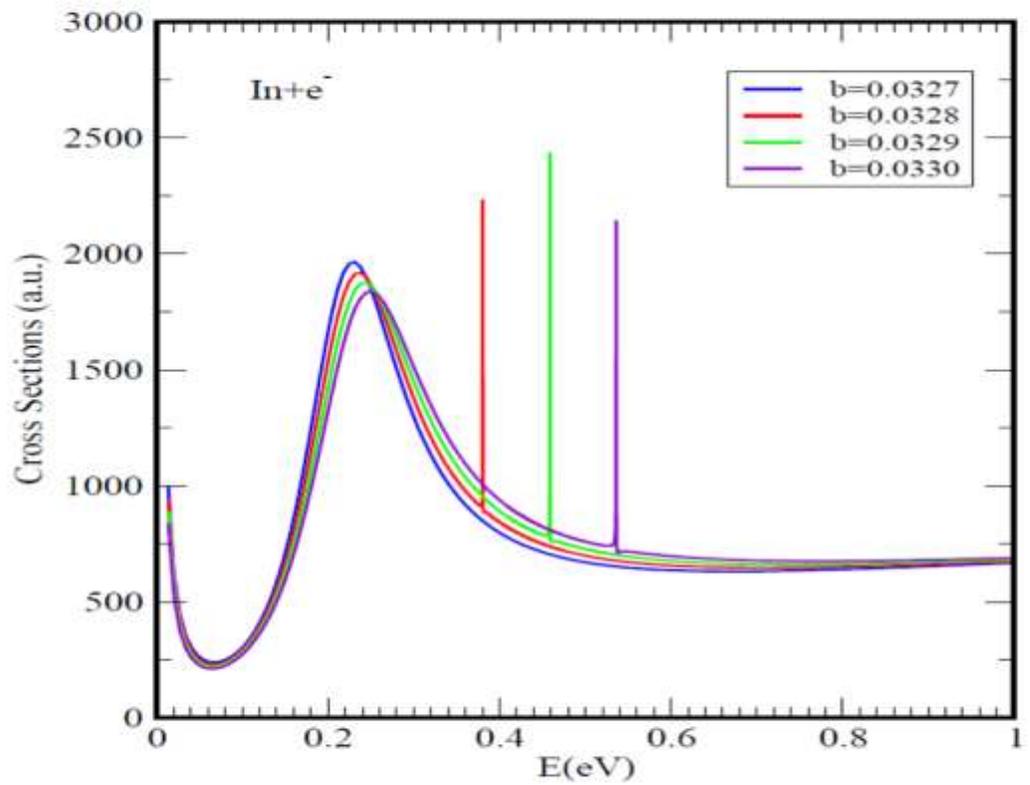

**Figure1:** Electron Elastic Scattering TCS (a.u.) versus E (eV) for atomic In, demonstrating the sensitivity of the electron affinity to the "b" parameter of the polarization potential.

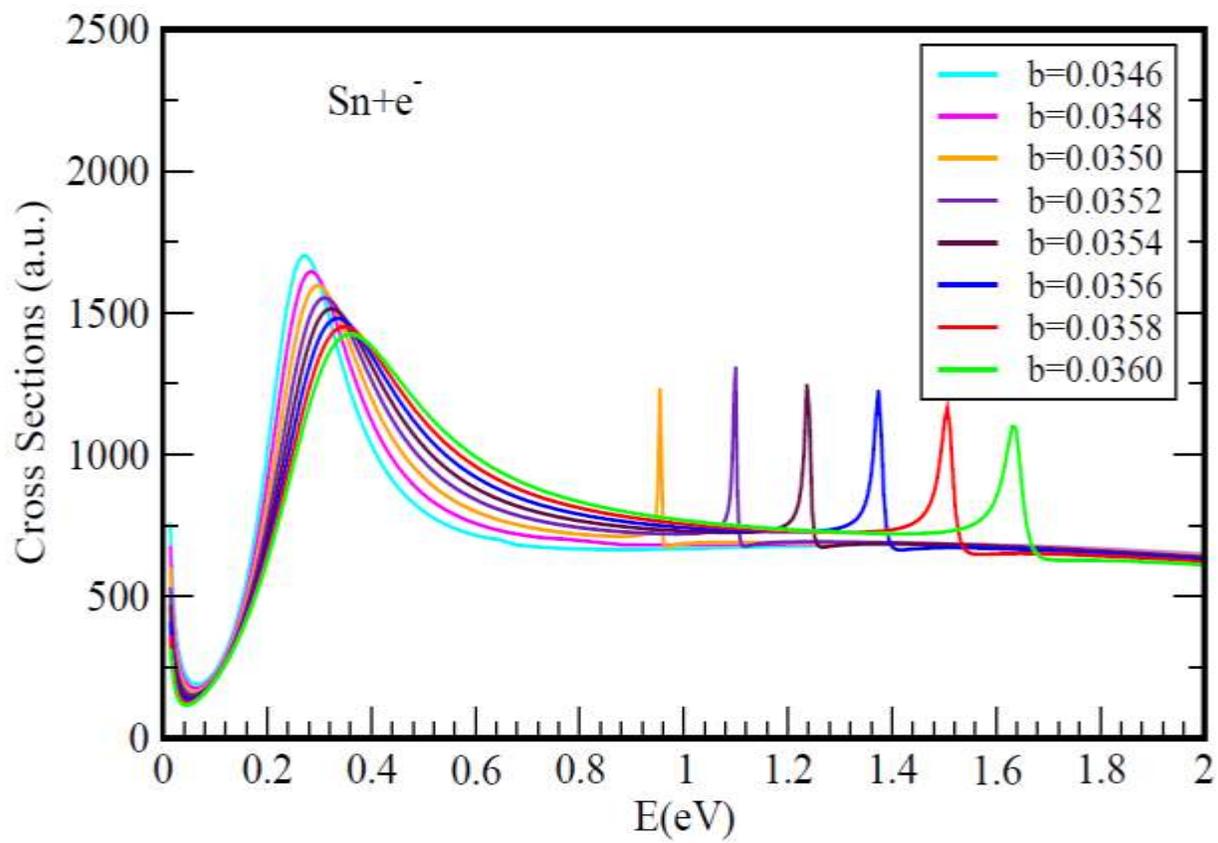

**Figure 2:** Electron Elastic Scattering TCS (a.u.) versus E (eV) for atomic Sn, demonstrating the sensitivity of the electron affinity to the "b" parameter of the polarization potential.

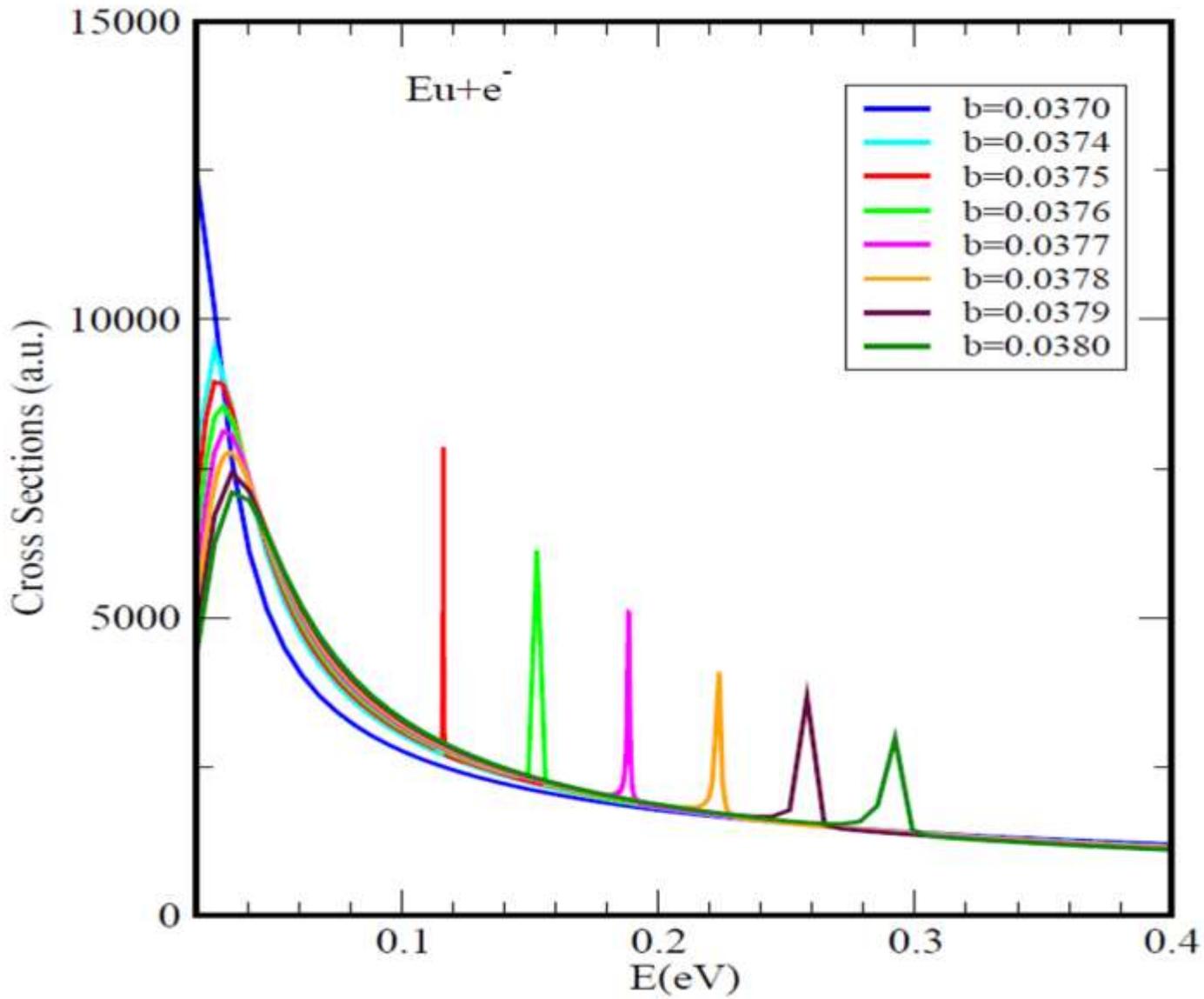

**Figure 3:** Electron Elastic Scattering TCS (a.u.) versus E (eV) for atomic Eu, demonstrating the sensitivity of the electron affinity to the "b" parameter of the polarization potential.

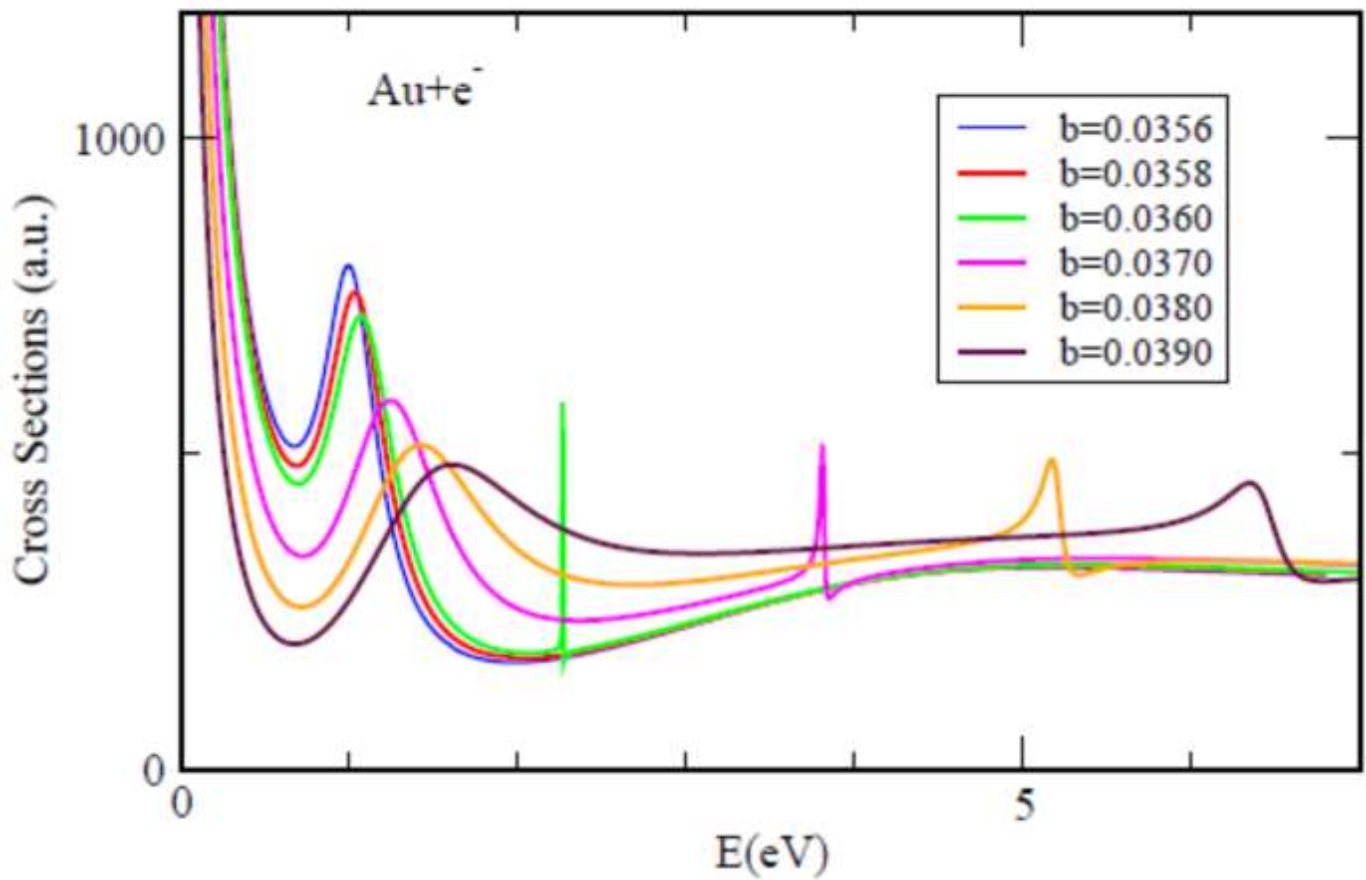

**Figure 4**: Electron Elastic Scattering TCS (a.u.) versus E (eV) for atomic Au, demonstrating the sensitivity of the electron affinity to the "b" parameter of the polarization potential.

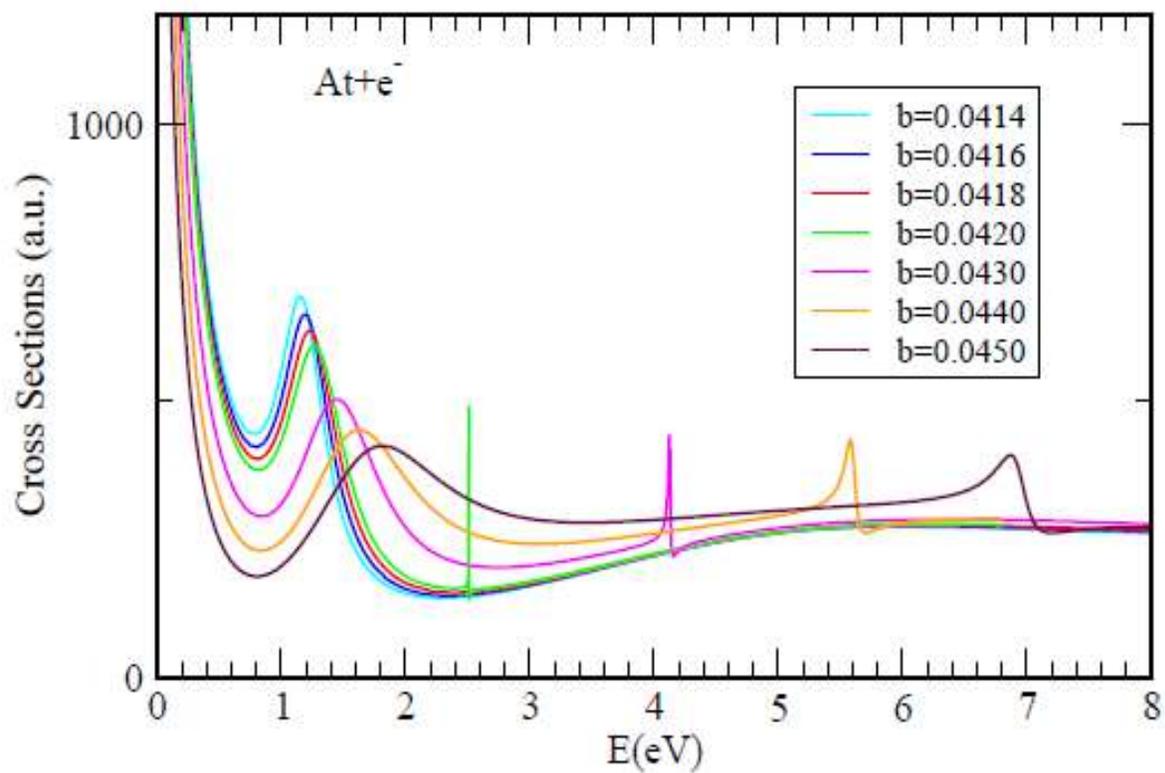

**Figure 5**: Electron Elastic Scattering TCS (a.u.) versus E (eV) for atomic At, demonstrating the sensitivity of the electron affinity to the "b" parameter of the polarization potential.